\documentstyle[12pt]{article}
\newcommand{\sect}[1]{\setcounter{equation}{0}\section{#1}\indent}
\renewcommand{\theequation}{\thesection.\arabic{equation}}
\renewcommand{\thefootnote}{\fnsymbol{footnote}}
\newcommand{\EQ}{\begin{equation}}
\newcommand{\EN}{\end{equation}}
\newcommand{\bea}{\begin{eqnarray}}
\newcommand{\ena}{\end{eqnarray}}
\newcommand{\vs}[1]{\vspace{#1 mm}}

\newcommand{\G}{\Gamma}
\renewcommand{\Im}{{\rm Im}\,}

\newcommand{\pa}{\partial}

\newcommand{\uda}{\nearrow \kern-1em \searrow}

\newcommand{\la}{\lambda}
\newcommand{\La}{\Lambda}

\makeatletter
\def\eqnarray{%
 \stepcounter{equation}%
 \let\@currentlabel=\theequation
 \global\@eqnswtrue
 \global\@eqcnt\z@
 \tabskip\@centering
 \let\\=\@eqncr
 $$\halign to \displaywidth\bgroup\@eqnsel\hskip\@centering
 $\displaystyle\tabskip\z@{##}$&\global\@eqcnt\@ne
 \hfil$\displaystyle{{}##{}}$\hfil
 &\global\@eqcnt\tw@$\displaystyle\tabskip\z@{##}$\hfil
 \tabskip\@centering&\llap{##}\tabskip\z@\cr}
\makeatother

\begin{document}

\begin{titlepage}
\setcounter{page}{0}
\begin{flushright}
EPHOU 96-005\\
September 1996\\
\end{flushright}

\vs{6}
\begin{center}
{\Large Periods and Prepotential of N=2 SU(2) Supersymmetric 
Yang-Mills 
Theory with Massive Hypermultiplets}

\vs{6}
{\large
Takahiro
Masuda
\footnote{e-mail address: masuda@phys.hokudai.ac.jp}
\\ and \\
Hisao Suzuki\footnote{e-mail address: hsuzuki@phys.hokudai.ac.jp}}\\
\vs{6}
{\em Department of Physics, \\
Hokkaido
University \\  Sapporo, Hokkaido 060 Japan} \\
\end{center}
\vs{6}

\centerline{{\bf{Abstract}}}
We derive a simple formula for the periods associated with   
the low energy effective action of $N=2$ supersymmetric $SU(2)$  
Yang-Mills theory with massive $N_f\le 3$ hypermultiplets. 
 This is given by evaluating explicitly the integral associated to the   
elliptic curve using various identities of hypergeometric functions.  
Following this formalism, 
 we can calculate the prepotential with massive hypermultiplets both in the weak coupling region and in the strong coupling region. 
In particular, we show how the Higgs field and its dual field are expressed as generalized 
hypergeometric functions when the theory has a conformal point.

\end{titlepage}
\newpage

\renewcommand{\thefootnote}{\arabic{footnote}}
\setcounter{footnote}{0}

\sect{Introduction} 

Since Seiberg and Witten discovered how to determine exactly 
the low-energy effective 
theory of $N=2$ supersymmetric $SU(2)$ Yang-Mills theory by 
using the elliptic curve\cite{SW},  
 many subsequent works have been made  on the basis of their analysis 
by extending the gauge group and 
by introducing matter hypermultiplets\cite{KLTY,APS,HO,Hanany,DS,AS}. 
The exact solution for 
the prepotential which controls the low energy effective action, 
can be obtained from the period 
integrals on the elliptic curve. Associated with singularities of 
the theories coming from the 
massless states, these curves for 
various kinds of $N=2$ supersymmetric Yang-Mills theories have been 
studied extensively\cite{HO}. 
Usual approach to obtain the period is to solve 
the differential equation which the periods obey, so called 
Pichard-Fuchs equation\cite{CDF,KLT}. When the theory is 
pure Yang-Mills with massless $N_f\le 3$ hypermultipletsthe with 
 gauge group $SU(2)$, this approach works 
successfully to solve the periods\cite{IY} because these theories 
have three singularity points if we use appropreate 
variables. Other more direct approach 
is known to be valid only in these cases\cite{Matone}. 
However when hypermultiplets are massive, the 
situation changes drastically; 
additional massless states appear in the theory 
and the number of singularities 
becomes more than three. Therefore, the Picard-Fuchs equation 
can no longer be solved by any special function 
 and the known solution is a perturbative 
solution in the weak coupling region\cite{Ohta}.

In this article we derive a simple formula for the periods 
from which we can obtain the prepotential both in the weak coupling region and in the strong coupling region; 
we can evaluate  the period integral of 
holomorphic one-form on the elliptic curve 
by using various identities of hypergeometric functions. 
As a result, the periods   
are represented as 
hypergeometric functions in terms of the function of $u$, $\La$ and masses, 
which are known from the form of the elliptic curve. We show that the resulting expression agrees with the results for massless case\cite{IY} and also have a power to handle the theories with conformal points.


\ 

\sect{Period Integrals}

We begin with reviewing some properties of the low-energy effective 
action of the $N=2$ supersymmetric $SU(2)$ QCD. 
In $N=1$ superfields formulation\cite{SW}, 
the theory contains chiral multiplets 
$\Phi^a$ and chiral field strength $W^a$ $(a=1,2,3)$ both in the adjoint 
representation of $SU(2)$, and chiral superfield $Q^i$ in ${\bf 2}$ and 
$\tilde{Q^i}$ $(i=1,\cdots, N_f)$ in $\bar{\bf 2}$ representation of 
$SU(2)$. In $N=2$ formulation $Q^i$ and $\tilde{Q^i}$ are called 
hypermultiplets. Along the flat direction, the scalar field $\phi$ of $\Phi$
 get vacuum expectation values which break $SU(2)$ to $U(1)$, so that 
the low-energy effective theory contains $U(1)$ vector multiplets
 $(A,W_{\alpha})$, where $A^i$ are $N=1$ chiral superfields and $W_{\alpha}$ 
are $N=1$ vector superfields. The quantum low-energy effective theory 
is characterized by effective Lagrangian $\cal L$ with 
the holomorphic function ${\cal F}(A)$ called prepotential,
\bea
{\cal L}={1\over 4\pi}\Im \left(\int d^2\theta 
d^2\bar{\theta} {\pa {\cal F}\over \pa A}\bar{A}
+{1\over 2}\int d^2\theta {\pa^2{\cal F}\over \pa A^2}W_{\alpha}W^{\alpha}
\right).
\ena
 The scalar component of $A$ is denoted by $a$, and $A_D={\pa {\cal F}
\over \pa A}$ which is dual to $A$ 
by $a_D$. The pair $(a_D,a)$ is a section of
$SL(2,{\bf Z})$ and is obtained as the period integrals of 
the elliptic curve parameterized by $u,\La$ and $m_i$ $(0\le i\le N_f)$, where
 $u=<$tr$\phi^2>$ is a gauge invariant moduli parameter,
$\La$ is a dynamical scale and $m_i$ are bare masses of
hypermultiplets. Once we know $a$ and $a_D$ as a holomorphic
function of $u$, we can calculate the prepotential ${\cal F}(a)$
by using the relation
\bea
a_D={\pa {\cal F}(a)\over \pa a}.
\ena

General elliptic curves of $SU(2)$ Yang-Mills theories with
massive $N_f \le 3$ hypermultiplets are\cite{HO} 
\bea
y^2=C^2(x)-G(x)
\ena
\begin{tabular}{lll}
$C(x)=x^2-u$,&$G(x)=\La^4$,&$(N_f=0)$\\
$C(x)=x^2-u$,&$G(x)=\La^3(x+m_1)$,&$(N_f=1)$\\
$C(x)=x^2-u+{\La^2\over 8}$,&$G(x)=\La^2(x+m_1)(x+m_2)$,&$(N_f=2)$\\
$C(x)=x^2-u+{\La\over 4}(x+{m_1+m_2+m_3\over 2})$,&$
G(x)=\La(x+m_1)(x+m_2)(x+m_3)$,&$(N_f=3)$\\
\end{tabular}

\

\noindent
These curves are formally denoted by
\bea
y^2=C^2(x)-G(x)=(x-e_1)(x-e_2)(x-e_3)(x-e_4),
\ena
where $e_1=e_4,\ e_2=e_3$ in the classical limit.
 In order to calculate the
prepotential, we consider $a$ and $a_D$ as the integrals of 
the meromorphic differential $\la$ over two 
independent cycles of these curves,
\bea
a&=&\oint_{\alpha}\la, \ \
a_{D}=\oint_{\beta}\la,\\
\la&=&{x\over 2\pi i}
\hbox{d} \ln\left({C(x)-y\over C(x)+y}\right).
\ena
where $\alpha$ cycle encloses $e_2$ and $e_3$, $\beta$ cycle encloses
$e_1$ and $e_3$, $\la$ is related
to the holomorphic one-form as
\bea
{\pa\la\over \pa u}={1\over 2\pi i}{dx\over y}
+d(*).
\ena
Since there are poles coming from mass parameters in the
 integrant of  $a$ and $a_D$,
we instead evaluate the period integrals of holomorphic one-form;
\bea
{\pa a\over \pa u}=\oint_{\alpha}{dx\over y},\ \
{\pa a_D\over \pa u}=\oint_{\beta}{dx\over y}.
\ena
First of all, we consider ${\pa a\over \pa u}$;
\begin{eqnarray}
{\pa a\over \pa u}={\sqrt 2\over 2\pi}
\int^{e_3}_{e_2}{dx\over y}={\sqrt 2\over 2\pi}
\int^{e_3}_{e_2}{dx\over \sqrt{(x-e_1)
(x-e_2)(x-e_3)(x-e_4)}},
\end{eqnarray}
where the normalization is fixed so as to be 
compatible with the asymptotic behavior of $a$ and $a_D$
in the weak coupling region
\bea
a&=&{\sqrt {2u}\over 2}+\cdots,\nonumber \\
a_D&=&i{4-N_f\over 2\pi}a\ln a+\cdots.\label{eq:asym}
\ena
After changing the variable  and using the integral representation of hypergeometric function; 
\bea
F(a,b;c;x)={\G(a)\G(b)\over \G(c)}\int_0^1 ds s^{b-1}(1-s)^{c-b-1}
(1-sx)^{-a}
\ena
where
\bea
F(a,b;c;z)=\sum_{n=0}^{\infty}{(a)_n(b)_n\over (c)_n}{z^n\over n!},
\ \ \ \ (a)_n={\G(a+n)\over \G(a)},
\ena
we obtain ${\pa a\over \pa u}$ as
\begin{eqnarray}
{\pa a\over \pa u}={\sqrt 2\over 2} (e_2-e_1)^{-1/2}(e_4-e_3)^{-1/2}
F\left({1\over 2},{1\over 2};1;z\right),\label{eq:a1}
\end{eqnarray}
where
\begin{eqnarray}
z={(e_1-e_4)(e_3-e_2)\over (e_2-e_1)(e_4-e_3)}.
\end{eqnarray}
Similarly we get the following expression for ${\pa a_D\over \pa u}$;
\begin{eqnarray}
{\pa a_{D}\over \pa u}&=&{\sqrt 2\over 2\pi}
\int^{e_3}_{e_1}{dy\over y}\nonumber \\
&=&{\sqrt 2\over 2}
 \left[(e_1-e_2)(e_4-e_3)\right]^{-1/2} 
F\left({1\over 2},{1\over 2},1;1-z\right).
\label{eq:aD1}
\end{eqnarray}
In this case $a_D$ is obtained as a hypergeometric function around 
$z=1$, so we have to do the analytic continuation which gives 
 the logarithmic asymptotic in the weak coupling region.

Since elliptic curves are not factorized in general,  
it is difficult to obtain their roots in a simple form. Even if we know the form of 
 roots, the variable $z$ in (\ref{eq:a1}) and (\ref{eq:aD1}) 
is very complicated in terms of $u$ in these 
representations. 
So we will transform the variable to the symmetric form 
 with respect to roots, by using the identity of the
hypergeometric functions, so that the 
new variable is given easily from the curve directly without 
knowing the form of roots.


\ 

\sect{Quadratic and cubic transformation}

\subsection{Quadratic transformation}

Before we treat a variety of $SU(2)$ Yang-Mills theory
with hypermultiplets, we consider the case 
where the elliptic curve is of the form  
\bea
y^2&=&(x^2+a_1x+b_1)(x^2+a_2x+b_2).\label{eq:curve1}
\ena
There are two possibilities that 
 $e_1$ and $e_2$ are roots of the 
first quadratic polynomial or $e_1$ and $e_4$ are .
 First of all, we consider the former case. 
If the variable of the hypergeometric function become 
 symmetric about $e_1$, $e_2$, and $e_3$, $e_4$, it is quite easy to
 read the variable from the form of this curve. To this end, we
 use the quadratic transformation\cite{HTF}
 for the hypergeometric functions to (\ref{eq:a1})
\bea
F\left(2a,2b;a+b+1/2;z\right)=F\left(
a,b;a+b+1/2;4z(1-z)\right),\label{eq:quad1}
\ena
where $a=b=1/4$, so that the new 
variable $z'=4z(1-z)$ of hypergeometric function is symmetric with respect to
 $e_1$, $e_2$ and $e_3$, $e_4$;
\bea
z'=4z(1-z)={(e_1-e_3)(e_2-e_4)(e_1-e_4)(e_3-e_2)\over
 (e_2-e_1)^2(e_4-e_3)^2},
\ena
and $z'$ can be easily expressed by $a_1,\ b_1,\ a_2,\ b_2$ as 
\bea
z'=-4{(b_1-b_2)^2-(b_1+b_2)a_1a_2+a_1^2b_2+a_2^2b_1
\over (a_1^2-4b_1)(a_2^2-4b_2)}.
\ena
Therefore, ${\pa a\over \pa u}$ can be written as
\bea
{\pa a\over \pa u}={\sqrt 2\over 2}
 [(e_2-e_1)(e_4-e_3)]^{-1/2} F\left({1\over 4},{1\over 4},1;z'\right).
\label{eq:a2}
\ena
Similarly for ${\pa a_D\over \pa u} $, after using the analytic continuation and the 
quadratic transformation, we get
\bea
{\pa a_D\over \pa u}={\sqrt 2\over 2}
 [(e_1-e_2)(e_4-e_3)]^{-1/2} \left[{6\ln 4\over 2\pi}
F\left({1\over 4},{1\over 4},1;z'\right)-{1\over \pi}
F^{*}\left({1\over 4},{1\over 4},1;z'\right)\right],
\label{eq:aD2}
\ena
where  
 $F^{*}(\alpha,\beta;1;z)$ is
another independent solution around $z=0$ of the differential equation 
 which $F(\alpha,\beta;1;z)$ obeys, which is expressed as
\bea
F^{*}\left(\alpha,\beta,1,z\right)=F(\alpha,&\beta&,1,z)\ln z
\nonumber \\
&+&\sum_{n=1}^{\infty}{(\alpha)_n (\beta)_n\over (n!)^2}
z^n\sum_{r=1}^{n-1}\left[
{1\over \alpha+n}+{1\over \beta+n}-{2\over n+1}\right].
\ena
Therefore, we obtain the general expression for ${\pa a\over \pa u}$ and 
${\pa a_D\over \pa u}$ in the 
weak coupling region valid in the case of the elliptic curve 
(\ref{eq:curve1}).
Notice that the quadratic transformation (\ref{eq:quad1}) is  valid if
$|z'|\le 1$. The region of $z$-plane
which satisfies this condition consists of
two parts; one is around $z=0$, one is around $z=1$. 
The region around $z=0$ corresponds to the
weak coupling region, and the regions around $z=1$ 
corresponds to the strong coupling region where 
monopole condensates. So we can
construct the formula valid in the strong coupling region by 
continuing the expression (\ref{eq:a1}) and (\ref{eq:aD1}) 
analytically to around $z=1$ and by applying the quadratic 
transformaion (\ref{eq:quad1}).

Similarly if we consider the latter case where $e_1$ and $e_3$ are roots of 
first quadratic polynomial of the curve (\ref{eq:curve1}), 
we have to do the transformation which make the variable symmetric 
about $e_1$, $e_4$ and $e_2,\ e_3$. Thus we use another 
quadratic transformaion\cite{HTF}
\bea
F\left(a,b;2b;z\right)=(1-z)^{-a/2}F\left({a\over 2},
b-{a\over 2};b+{1\over 2};{z^2\over 4(1-z)}\right),\label{eq:quad2}
\ena
where $a=1/2$. The new variable $\tilde{z}'=z^2/4(1-z)$ 
is symmetric about $e_1,e_4$ and $e_2,e_3$ 
as follows;
\bea
\tilde{z}'&=&{z^2\over 4(1-z)}={(e_1-e_4)^2(e_3-e_2)^2\over
 4(e_2-e_1)(e_4-e_3)(e_1-e_3)(e_4-e_2)}\nonumber \\
&=&-{(a_1^2-4b_1)(a_2^2-4b_2)\over 
4[(b_1-b_2)^2-(b_1+b_2)a_1a_2+a_1^2b_2+a_2^2b_1]}
\ena
By applying this transformation to (\ref{eq:a1}) and (\ref{eq:aD1}), 
we get ${\pa a\over \pa u},\ {\pa a_D\over \pa u}$ as
\bea
{\pa a\over \pa u}&=&{\sqrt 2\over 2}
 [(e_3-e_1)(e_4-e_2)(e_2-e_1)(e_4-e_3)]^{-1/4} 
F\left({1\over 4},{1\over 4},1;\tilde{z}'\right),
\label{eq:a3}\\
{\pa a_D\over \pa u}&=&i{\sqrt 2\over 2}
 [(e_1-e_3)(e_4-e_2)(e_2-e_1)(e_4-e_3)]^{-1/4} \nonumber \\
& & \hspace{1cm}\times \left[{3\ln 4-i\pi\over 2\pi}
F\left({1\over 4},{1\over 4},1;\tilde{z}'\right)-{1\over 2\pi}
F^{*}\left({1\over 4},{1\over 4},1;\tilde{z}'\right)\right].
\label{eq:aD3}
\ena
In both cases we can read the variable directly from the coefficients 
of the curve. 

In the next subsection, we generalize the formalism of this subsection 
to all kinds of $SU(2)$ Yang-Mills theory
 with massive $N_f\le 3$ hypermultiplets. 

\subsection{Cubic transformation}

We denote the curve as
\bea
y^2=x^4+ax^3+bx^2+cx+d.
\ena
In general,  
the variable of the hypergeometric function is still very complicated even after the quadratic transformation. 
So in addition to the quadratic transformation, we must use 
the the following cubic transformation\cite{HTF}
 subsequently
\bea
F\left(3a,a+{1\over 6};4a+{2\over 3};z'\right)=
\left(1-{z'\over 4}\right)^{-3a}F\left(a,a+{1\over 3};2a+{5\over 6};
-27 {z'^2\over (z'-4)^3}\right),\label{eq:cub}
\ena
or 
\bea
F\left(3a,{1\over 3}-a;2a+{5\over 6};\tilde{z}'\right)=
(1-4\tilde{z}')^{-3a}F\left(a,a+{1\over 3};2a+{5\over 6};
{27\tilde{z}'\over (4\tilde{z}'-1)^3}\right),
\label{eq:cub2}
\ena
where $a=1/12$, 
so that the new variable $z''=-27z'^2/(z'-4)^3=27\tilde{z}'/
(4\tilde{z}'-1)^3$ 
become completely symmetric in $e_i$. 
Notice that $z''$ is represented by coefficients of the elliptic curve 
\bea
z''=-27 {z'^2\over (z'-4)^3}=27 {\tilde{z}'\over (
4\tilde{z}'-1)^3}={27 z^2 (1-z)^2\over 4 (z^2-z+1)^3}
=-{27\Delta\over 4D^3},
\ena
where $\Delta$ is the discriminant of the elliptic curve 
\bea
\Delta&=&\prod_{i<j}(e_i-e_j)^2\nonumber \\
&=&-[27 a^4 d^2+a^3c(4 c^2-18bd)+ac(-18bc^2+80 b^2 d+192 d^2)\nonumber \\
& &\ \ + 
a^2(-b^2 c^2+4b^3d+6c^2d-144bd^2)+4b^3c^2+27c^4\\
& & \ \ \ \ -16b^4d-144bc^2d+
128b^2d^2-256d^3],\nonumber
\ena
and $D$ is given by
\bea
D=\sum_{i<j}{1\over 2}(e_i-e_j)^2=-b^2+3ac-12d.
\ena
Applying (\ref{eq:cub}) to (\ref{eq:a2}) or (\ref{eq:cub2}) to (\ref{eq:a3}), 
without knowing precise forms of $e_i$  
we obtain a general expression for ${\pa a\over \pa u}$ 
in the weak coupling region 
valid even in the theory with massive $N_f\le 3$ hypermultiplets,
\bea
{\pa a\over \pa u}={\sqrt 2\over 2}(-D)^{-1/4}F\left(
{1\over 12},{5\over 12};1;-{27\Delta\over 4D^3}\right).\label{eq:a4}
\ena 
Similarly, after the analytic continuation and 
quadratic and cubic transformations,
 we obtain an expression for ${\pa 
a_D\over \pa u}$ as 
\bea
{\pa a_{D}\over \pa u}=i{\sqrt 2\over 2}(-D)^{-1/4}
\left[ {3\over 2\pi}\ln 12 \,\right. &F&\left({1\over 12},{5\over 12},1,
-{27\Delta\over 4D^3}\right) \nonumber \\
&-&\left. {1\over 2\pi}F^{*}\left(
{1\over 12},{5\over 12};1;-{27\Delta\over 4D^3}\right)\right].
\label{eq:aD4}
\ena
For the consistency check, we consider the asymptotic 
behavior in the weak coupling region $u\rightarrow \infty$,
\bea
\Delta&=&(-1)^{N_f}256u^{N_f+2}\La^{2(4-N_f)}+\cdots, \nonumber \\
D&=&-16u^2+\cdots,\\
-{27\over 4}{\Delta\over D^3}&=&{27(-1)^{N_f}\over 64}\left({\La^2\over 
u}\right)^{4-N_f}+\cdots.\nonumber
\ena
Thus we have
\bea
{\pa a\over \pa u}&=&{\sqrt 2\over 4\sqrt u}+\cdots,
\nonumber \\
{\pa a_D\over \pa u}&=&i{\sqrt 2\over 4\sqrt u}{4-N_f\over 2\pi}
\ln\left({\La^2\over u}\right)+\cdots,
\ena
which is compatible with (\ref{eq:asym}).

The formula (\ref{eq:a4}) and (\ref{eq:aD4}) are useful in the case where 
we cannot obtain any simple expression of roots, whereas 
 (\ref{eq:a2}) and (\ref{eq:aD2}) or (\ref{eq:a3}) and (\ref{eq:aD3}) can be 
used when we have a factorized form for $y^2$ as (\ref{eq:curve1}).

Next we consider the periods in the strong coupling region.
The quadratic and cubic transformation are valid if 
$|z''|\le 1$. The region of $z$-plane 
which satisfies this condition consists of 
three parts; one is around $z=0$ , one is around $z=1$ and the last is 
around $z=\infty$. The region around $z=0$ corresponds to the 
weak coupling region, and the region around $z=1$  
corresponds to the strong coupling region where the monopoles condensate and $z=\infty$ is the dyonic point. So we can 
construct the formula valid in the strong coupling region by 
analytic continuation to around $z=1$ or $z=\infty$ and by 
using the quadratic and cubic transformation subsequently.
For example, the formula around the strong coupling region $z=1$ is given by 
\bea
{\pa a\over \pa u}&=&{\sqrt 2\over 2}(-D)^{-1/4}
\left[ {3\over 2\pi}\ln 12 \,\right. F\left({1\over 12},{5\over 12},1,
-{27\Delta\over 4D^3}\right) \nonumber \\
& &\hspace{4cm}-\left. {1\over 2\pi}F^{*}\left(
{1\over 12},{5\over 12};1;-{27\Delta\over 4D^3}\right)\right],\\
\label{eq:a5}
{\pa a_D\over \pa u}&=&i{\sqrt 2\over 2}(-D)^{-1/4}F\left(
{1\over 12},{5\over 12};1;-{27\Delta\over 4D^3}\right).\label{eq:aD5}
\ena
The expression (\ref{eq:a4}), (\ref{eq:aD4}) and 
(\ref{eq:a5}), (\ref{eq:aD5}) show a manifest duality of the periods.

Notice that the ratio of two period integrals is 
the coupling constant of the theory,
\bea
\tau={\pa^2 {\cal F}\over \pa a^2}={\pa a_D\over \pa a}=
\left.{\pa a_D\over \pa u}\right/{\pa a\over \pa u}=
{iF(1/2,1/2,1,1-z)\over F(1/2,1/2,1,z)}.
\ena
Though $z$ is not invariant under the modular transformation of $\tau$,
 the argument $z''=-27\Delta/4D^3=27z^2(1-z)^2/(z^2-z+1)^3$ is
 invariant completely. As a matter of fact, this variable can be written by the absolute invariant form $j(\tau)$ as $z''=1/j(\tau)$. Therefore it is quite 
natural to represent the period in terms of $z''$.


\ 

\sect{Examples}

In this section we calculate the period, $a$, $a_D$ and the 
prepotential of a variety of supersymmetric 
$SU(2)$ Yang-Mills theory with massive hypermultiplets as 
examples of our formula. 
For consistency check we also consider massless case. 
 Moreover we consider the cases where the theory has 
conformal points.

\subsection{$N_f=1$ theory} 

We 
consider the theory with a matter hypermultiplet whose curve is given by
 \bea
y^2=(x^2-u)^2-\La^3(x+m),
\ena
from which $\Delta$ and $D$ is obtained as
\bea
\Delta&=&-\La^6(256u^3-256u^2m^2-288um\La^3+256m^3\La^3+27\La^6),\\
D&=&-16u^2+12m\La^3.
\ena
Substituting these to (\ref{eq:a4}) and (\ref{eq:aD4}), we 
can obtain 
$a,\ a_D$, by expanding (\ref{eq:a4}) and (\ref{eq:aD4}) at $u=\infty$ and  
integrating with respect to $u$. Representing $u$ in terms of 
 $a$ inversely, and substituting $u$ to $a_D$, and finally integrating $a_D$
with respect to $a$, we can get the prepotencial in the 
weak coupling region as
\bea
{\cal F}(\tilde{a})&=&i{\tilde{a}^2\over \pi}\left
[{3\over 4}\ln \left({\tilde{a}^2\over \La^2}\right)+{3\over 4}\left(
-3+4\ln 2-i\pi\right)-{\sqrt 2\pi\over 2i\tilde{a}}
(n'm)\right.\nonumber \\
& & \left.-\ln \left({\tilde{a}\over \La}\right){m^2\over 4\tilde{a}^2}
+\sum_{i=2}^{\infty}{\cal F}_i\tilde{a}^{-2i}\right].
\label{eq:pre2}
\ena
where we introduce $\tilde{a}$ subtracted mass residues
from $a$. These ${\cal F}_i$
agree with the perturbative result up to the orders cited
in \cite{Ohta}. In principle we can calculate ${\cal F}_i$
to arbitrary order in our formalism. Quite similarly, we can obtain the prepotential in the strong coupling region.

 To compare to the periods for massless case where the explicit form is known
  by solving the Picard-Fuchs equation\cite{IY}, we start with our 
expression for the massless theory,
\bea
\Delta=-\La^6(256u^3+27\La^6),\ 
D=-16u^2,\ z''=-{27\over 4}{\La^6(256u^3+27\La^6)\over 16^3u^6}.
\ena
If we set $w=-27\La^6/256u^3$ then $z''=4w(1-w)$, thus using 
the quadratic transformations \\(\ref{eq:quad1}), 
we get the expression for the massless case,
\bea
{\pa a\over \pa u}&=&{\sqrt 2\over 2}{1\over 2\sqrt u}F\left(
{1\over 6},{5\over 6},1;w\right),\nonumber \\
{\pa a_D\over \pa u}&=&i{\sqrt 2\over 2}{1\over 2\sqrt u}
\left[{3\ln 3+2\ln 4\over 2\pi}F\left({1\over 6},{5\over 6}
,1,w\right)-{1\over 2\pi}F^{*}\left({1\over 6},{5\over 6}
,1,w\right) \right].
\ena
Integrating with respect to $u$, we can 
get the expression given by Ito and Yang\cite{IY}.
The expression around the strong coupling region can be obtained from $(3.22)$ and $(3.22)$ by using the identity $(3.2)$ for $w=1-{27\Lambda^6\over 256u^3}$.
In general when $m\ne 0$, because of the
singularity comming from additional massless states\cite{SW},
we cannot represent $a$ and $a_D$ as any
special functions by integrating the
expression for ${\pa a\over \pa u}$ and ${\pa a_D\over \pa u}$.
However if masses take critical values with which the number of the
singularity goes down to the same number
as in the massless case, that is three,
 $a$ and $a_D$ seem to be expressed by special functions.
 The number of the singularity is the number of the root of the 
equation $\Delta=0$ plus one, which is the singularity at $u=\infty$.
Since $\Delta$ is third order polynomial in terms of $u$ in $N_f=1$ theory,
$\Delta=0$ must have one double root when the mass takes 
critical value. 
This condition is satisfied if $m=3/4\La$ where the parameters of the periods are given by 
\bea
\Delta&=&-\La^6(16u+15\La^2)(4u-3\La^2)^2,\ \ 
D=-(4u+3\La)(4u-3\La),\\
z''&=&-{27\over 4}{\La^6(16u+15\La^2)\over (4u+3\La^2)^3(4u-3\La^2)}.
\ena
Such factorization of $\Delta$ means that 
theory has a conformal point $u=3\La^2/4$ where the curve become
\cite{APSW,EHIY} 
\bea
y^2=\left(x+{\La\over 2}\right)^3\left(x-{3\La\over 2}\right).
\ena
If we set 
\bea
w={27\La^2\over 16u+15\La^2},
\ena
then $z''=-64w^3/(w-9)^3(w-1)$. 
In order to obtain $a$ and $a_D$ 
we need the quartic transformation which makes the variable 
simple enuogh. We can prove the following  transformation of fourth order; 
\bea
F\left({1\over 12},{5\over 12},1,-{64w^3\over (w-9)^3(w-1)}\right)=
\left(1-{w\over 9}\right)^{1/4}(1-w)^{1/12}F\left({1\over 3},
{1\over 3},1,w\right).
\ena
Using this identity and the identity
\bea
F(a,b;c;z)=(1-z)^{-a}F(a,c-b;c;z/(z-1)),\label{eq:ide}
\ena
we get ${\pa a\over \pa u},\ 
{\pa a_D\over \pa u}$ 
\bea
{\pa a\over \pa u}&=&{\sqrt2 \over 8}\left(-27\La^2\right)^{1\over 2}
y^{1/2}F\left({1\over 3},
{2\over 3},1,y\right) \label{eq:paa2}\\
{\pa a_D\over \pa u}&=&i{\sqrt 2\over 8}\left(-27\La^2\right)^{1\over 2}
y^{1/2}\left[{\left(3\ln 3-i\pi\right)\over 2\pi}
F\left({1\over 3},{2\over 3},1,y\right)\right.\nonumber \\
& &\hspace{7cm}-\left.{3\over 2\pi}F^{*}
\left({1\over 3},{2\over 3},1,y\right)\right],\label{eq:paaD2}
\ena
where
\bea
y={27\La^2\over -16u+12\La^2}.
\ena
Integrating with respect $u$, 
we get $a$ and $a_D$ in the weak coupling region as
\bea
a&=&-{i\sqrt 2\over 8}3\sqrt{3}\La
 {y^{-{1\over 2}}}_3F_2\left({1\over 3},{2\over 3},-{1\over 2}
;1,{1\over 2};y\right),\label{eq:formula1}\\
a_D&=&+{\sqrt 2\over 8}3\sqrt{3}\La
 y^{-{1\over 2}}\left[
{3(3\ln 3-i\pi-2)\over 2\pi}\, _3F_2\left({1\over 3},{2\over 3},-{1\over 2}
;1,{1\over 2};y\right)\right.\nonumber \\
& &\hspace{4cm}
\left.-{3\over 2\pi}\, _3F^{*}_2\left({1\over 3},{2\over 3},-{1\over 2}
;1,{1\over 2};y\right)\right],\label{eq:formula1D}
\ena
where $_3F_2(a,b,c;1,d,y)$ is the generalised hypergeometric function\cite{HTF}
\bea
_3F_2(a,b,c;1,d;y)
=\sum_{n=0}^{\infty}{(a)_n (b)_n 
(c)_n\over (d)_n (n!)^2}y^n,
\ena
and 
\bea 
_3F_2^{*}(a,b,c;1,d;y)
&=& _3F_2(a,b,c;1,d;y)\ln y\nonumber \\
&+&\sum_{n=0}^{\infty}{(a)_n (b)_n
(c)_n\over (d)_n (n!)^2}y^n\\
& &\times \sum_{r=0}^{n-1}\left[{1\over a+r}
+{1\over b+r}+{1\over c+r}-
{1\over d+r}-{2\over 1+r}\right],\nonumber
\ena
is other independent solutions of 
a generalized hypergeometric equation\cite{HTF} around $y=0$;
\bea
y^2(1-y){d^3 F\over dy^3}
&+&\{(d+2)y-(3+a+b+c)y^2\}{d^2 F\over dy^2}
\nonumber \\
+&\{&d-(1+a+b+c+ab+bc+ca)y\}{dF\over dy}-abc F=0,
\ena
which $_3F_2(a,b,c;1,d;y)$ obeys. Notice that Picard-Fuchs equation of 
$N_f=1$ theory reduces to this equation when the theory has the conformal 
point. This equation has three regular singularities at $y=0,1,\infty$. 
This is the reason why $a$ and $a_D$ are expressed by the special functions. 

In order to obtain the expression around the conformal point $u=3\La/4$ 
from the expression (\ref{eq:formula1}) and (\ref{eq:formula1D}), 
we have to perform the analytic continuation from  
the weak coupling region. After that, the expressions 
for $a$ and $a_D$ 
contain no logarithmic terms,
\bea
a&=&-{i\sqrt 2\over 8}3\sqrt{3}\La
y^{-1/2}\left[
{6\over 5}{\Gamma({1\over 3})\over \Gamma({2\over 3})\Gamma({2\over 3})}
(-y)^{-1/3}\, _3F_2\left({1\over 3},{1\over 3},{5\over 6};{2\over 3}
,{11\over 6};{1\over y}\right)
\right. \nonumber \\
& &\ \ +\left.{6\over 7}{\Gamma(-{1\over 3})\over \Gamma({1\over 3})
\Gamma({1\over 3})}
(-y)^{-2/3}\, _3F_2\left({2\over 3},{2\over 3},{7\over 6};{4\over 3}
,{13\over 6};{1\over y}\right)
\right],\\
a_D&=&{\sqrt 3\over 2}
{\sqrt 2\over 8}3\sqrt{3}\La
y^{-1/2}\left[
{6\over 5}{\Gamma({1\over 3})\over \Gamma({2\over 3})\Gamma({2\over 3})}
(-y)^{-1/3}\, _3F_2\left({1\over 3},{1\over 3},{5\over 6};{2\over 3}
,{11\over 6};{1\over y}\right)
\right. \nonumber \\
& &\ \ -\left.{6\over 7}{\Gamma(-{1\over 3})\over \Gamma({1\over 3})
\Gamma({1\over 3})}
(-y)^{-2/3}\, _3F_2\left({2\over 3},{2\over 3},{7\over 6};{4\over 3}
,{13\over 6};{1\over y}\right)
\right],\ena
where $1/y= -16u/27\Lambda^{2} + 4/9$. Thus the coupling constant $\tau$ which is the ratio of ${\pa a_D\over \pa u}$ 
and ${\pa a \over \pa u}$
 has no logarithmic term, and the beta function on this conformal point 
 vanishes. 

Here we pause to discuss an interesting  relation  
between the moduli space of this theory and the moduli space of 
2-D N=2 superconformal field theory with central charge $c=3$. 
Consider the 
complex projective space ${\bf P}^2$ 
with homogeneous coordinates $[x_0,x_1,x_2]$ and define the 
hypersurface $X$ by the equation
\bea
f=x_0^3+x_1^3+x_2^3-3\psi x_0x_1x_2=0.
\ena
Moduli space of the theory with $c=3$ is described 
by $\tau$, the ratio of two independent period integrals 
of holomorphic one-form $\Omega$ over the cycle on $X$,
\bea
\tau=\left.\int_{\gamma}\Omega\right/\int_{\gamma'}\Omega.
\ena
It is known that this period satisfy the Picard-Fuchs equation
 which reduce to 
\bea
\left(z{d\over dz}\right)^2f(z)-z\left(z{d\over dz}+{1\over 3}\right)
\left(z{d\over dz}+{2\over 3}\right)f(z)=0,
\ena
where $z=\psi^{-3}$. This is a hypergeometric differential equation and 
$f(z)$ is obtained as a linear combination
 of $F(1/3,2/3;1;z)$ and $F^{*}(1/3,2/3;1;z)$.
 By comparing this solution  to (\ref{eq:paa2}) and 
(\ref{eq:paaD2}), we deduce 
an identification $\psi^3 =  
-16u/27\La^2+ 4/9$, and that 
the conformal point $(u=3\La^2/4)$ of 4-D $SU(2)$ $N_f=1$ super QCD 
 corresponds to the Landau-Ginzburg point $(\psi=0)$ of 
2-D SCFT with $c=3$. It seems interesting to use this identification to investigate the theory at the conformal fixed point.


\subsection{$N_f=2$ theory}

We consider the theory with $N_f=2,\ m_1=m_2=m$ whose curve and descriminant are given by
\bea
y^2&=&\left(x^2-u+{\La^2\over 8}\right)-\La^2(x+m)^2\nonumber \\
&=&\left(x^2-\La x-\La m-u+{\La^2\over 8}\right)
\left(x^2+\La x+\La m-u+{\La^2\over 8}\right) \nonumber \\
\Delta&=&{\La^2\over 16}\left(8u-8m^2-\La^2\right)^2\left(8u+8\La m+\La^2\right)
\left(8u-8\La m+\La^2\right).
\ena
In this case, we can use the formula of section 3.1 because of the factorized form of the curve. 
Reading $z'$, $e_2-e_1$, $e_4-e_3$ from the coefficients of the curve, 
\bea
(e_2-e_1)^2&=&4u+4\La m+{\La^2\over 2},\nonumber \\
(e_4-e_3)^2&=&4u-4\La m+{\La^2\over 2}, \\
z'&=&{\La^2 (u-m^2-{\La^2\over 8})\over (u+\La m+{\La^2\over 8})(
u-\La m+{\La^2\over 8})}.\nonumber
\ena
Substituting these to (\ref{eq:a2}) and (\ref{eq:aD2}), we 
can obtain $a$ and $a_D$ after expansion around 
$u=\infty$ and integration with respect $u$. The prepotential in the 
weak coupling region is
\bea
{\cal F}(\tilde{a})&=&i{\tilde{a}^2\over \pi}\left
[{1\over 2}\ln \left({\tilde{a}^2\over \La^2}\right)+\left(
-1+{i\pi \over 2}+{5\ln 2\over 2}\right)-{\sqrt{2}\pi\over 2i\tilde{a}}
(n'm)\right.\nonumber \\
& & \left.-\ln \left({\tilde{a}\over \La}\right){m^2\over 2\tilde{a}^2}
+\sum_{i=2}^{\infty}{\cal F}_i\tilde{a}^{-2i}\right].
\label{eq:pre1}\ena
These ${\cal F}_i$ agree with the result up to known orders \cite{Ohta}.
We can calculate the prepotential in $m_1\ne m_2$ case. 
 Let us compare the results of $a$ and $a_D$ for massless case to the previous results\cite{IY}. 
The variable $z$ can be written in the form  $z'=4w(1-w)$ if we set $x=\La^2/8u$ and $w=2x/(x+1)$. Therefore,  we transform
 $z'$ to $w$ by using the identity (\ref{eq:quad1}) and 
 $w$ to $x^2$ by using the identity
\bea
F(a,b;2b;w)=\left(1-{z\over 2}\right)^{-a}
F\left({a\over 2},{1\over 2}+{a\over 2};b+{1\over 2};
{w^2\over (w-2)^2}\right),
\ena
where $a=b=1/2$ and $w^2/(w-2)^2=x^2$. Thus we get the expression for 
the massless case;
\bea
{\pa a\over \pa u}&=&{\sqrt 2\over 2}{1\over 2\sqrt u}F\left(
{1\over 4},{3\over 4},1;x^2\right),\nonumber \\
{\pa a_D\over \pa u}&=&i{\sqrt 2\over 2}{1\over 2\sqrt u}
\left[{3\ln 4\over 2\pi}F\left({1\over 4},{3\over 4}
,1,x^2\right)-{1\over 2\pi}F^{*}\left({1\over 4},{3\over 4}
,1,x^2\right) \right].
\ena
Integrating with respect to $u$, we can recover the previous 
result for $a$ and $a_D$ \cite{IY}. 

Next we consider the case where  
 the same factor appears in the denominator and the numerator 
of $z'$. This is satisfied if 
$m=\La/2$. In this case the theory has conformal point\cite{APSW,EHIY} at 
$u=3\La^2/8$ where the elliptic curve is factorized as
\bea
y^2=\left(x+{\La\over 2}\right)^3\left(x-{3\La\over 2}\right).
\ena
Main defferenece between the massless theory and the massive theory is 
the existence of this conformal point. Usual massive $N_f=2$ theory has 
five singularity points where additional two singularity points are coming 
from the two bare mass parameter. 
In this subsection we set $m_1=m_2$, so the number 
of the singularity is four. When the theory has conformal point, 
two of four singularity point coincide and this point becomes a conformal 
point\cite{APSW,EHIY}. On the other hand, in our representation since 
the pole and the zero of the variable $z'$  
of the hypergeometric function correspond to the singularity points, 
the theory has 
a conformal point when the pole and the zero of $z'$ 
coincide and this pole becomes a conformal point.

To obtain $a$ and $a_D$ of this theory, 
we substitute $m=\La/2$ to (\ref{eq:a2}) and (\ref{eq:aD2}), and  
use the identity (\ref{eq:ide}), 
 we obtain ${\pa a\over \pa u},\ {\pa a_D\over \pa u}$
\bea
{\pa a\over \pa u}&=&{\sqrt2 \over 2}(-\La^2)^{-1/2}
y^{1/2}F\left({1\over 4},
{3\over 4},1,y\right)\label{eq:paa} \\
{\pa a_D\over \pa u}&=&i{\sqrt 2\over 2}(-\La^2)^{-1/2}
y^{1/2} \left[{\left(6\ln 4-i\pi\right)\over 2\pi}
F\left({1\over 4},{3\over 4},1,y\right)\right.\\
& &\hspace{7cm}\left.-{1\over \pi}F^{*}
\left({1\over 4},{3\over 4},1,y\right)\right],\label{eq:paaD}
\ena
where 
\bea
y={8\La^2\over -8u+3\La^2}.
\ena
Integrate with respect to $u$, we get $a$ and $a_D$ in the weak coupling 
region as
\bea
a&=&-{\sqrt 2\over 2}(-1)^{1\over 2}\La {y^{-{1\over 2}}}
_3F_2\left({1\over 4},{3\over 4},-{1\over 2};
1,{1\over 2};y\right),\label{eq:aw}\\
a_D&=&-i{\sqrt 2\over 2}(-1)^{1\over 2}\La y^{-{1\over 2}}
\left[{(6\ln 4-i\pi-4)\over 2\pi}
\, _3F_2\left({1\over 4},{3\over 4},-{1\over 2};
1,{1\over 2};y\right)\right.\nonumber \\
& &\hspace{5cm} \left.-{1\over \pi}\, 
_3F^{*}_2\left({1\over 4},{3\over 4},-{1\over 2};
1,{1\over 2};y\right)\right].\label{eq:aDw}
\ena

As $N_f=1$ theory, after the use of the analytic continuation from 
the weak coupling region, we obtain $a$ and $a_D$
around conformal point $u=3\La^2/8$ as follows; 
\bea
a&=&-{\sqrt 2\over 2}(-1)^{1\over 2}\La
y^{-1/2}\left[
{4\over 3}{\Gamma({1\over 2})\over \Gamma({3\over 4})\Gamma({3\over 4})}
(-y)^{-1/4}\, _3F_2\left({1\over 4},{1\over 4},{3\over 4};{1\over 2}
,{7\over 4};{1\over y}\right)
\right. \nonumber \\
& &\ \ +\left.{4\over 5}{\Gamma(-{1\over 2})\over \Gamma({1\over 4})
\Gamma({1\over 4})}
(-y)^{-3/4}\, _3F_2\left({3\over 4},{3\over 4},{5\over 4};{3\over 2}
,{9\over 4};{1\over y}\right)
\right],\\
a_D&=&-i{\sqrt 2\over 2}(-1)^{1\over 2}\La
y^{-1/2}\left[{4\over 3}
{\Gamma({1\over 2})\over \Gamma({3\over 4})\Gamma({3\over 4})}
(-y)^{-1/4}\, _3F_2\left({1\over 4},{1\over 4},{3\over 4}
;{1\over 2},{7\over 4};{1\over y}\right)
\right. \nonumber \\
& &\ \ -\left.{4\over 5}{\Gamma(-{1\over 2})\over \Gamma({1\over 4})
\Gamma({1\over 4})}
(-y)^{-3/4}\, _3F_2\left({3\over 4},{3\over 4},{5\over 4};{3\over 2}
,{9\over 4};{1\over y}\right)
\right].\ena

\subsection{$N_f=3$ theory}

As in $N_f=1$ theory we read $\Delta$ and $D$ from the curve 
although they are much 
more complicated because of many bare mass parameter $m_i$. 
After substituting these to (\ref
{eq:a4}) and (\ref{eq:aD4}) and using similar manner as $N_f=1,2$ case, we 
get the prepotentical in the weak coupling region as
\bea
{\cal F}(\tilde{a})&=&{i\tilde{a}^2\over \pi}\left[
{1\over 4}\ln \left({\tilde{a}^2\over \La^2}\right)+
{1\over 4}(9\ln 2-2-\pi i)-{\sqrt 2 \pi\over 4i\tilde{a}}
\sum_{i=1}^{3}n'_im_i\right.\nonumber \\
& & \left. -{1\over 4\tilde{a}^2}\ln\left({\tilde{a}\over \La}\right)
\sum_{i=1}^3 m_i^2+\sum_{i=2}^{\infty}{\cal F}_i\tilde{a}^{-2i}\right].
\label{eq:pre3}
\ena
These ${\cal F}_i $ agree the result up to known orders \cite{Ohta}.

Let us consider massless case where $\Delta$ and 
$D$ are given by
\bea
\Delta&=&-\La^2u^4(-\La^2+256u),\ \ D={-\La^4+256\La^2u-4096u^2\over 256},
\nonumber \\
z''&=&{27(256)^3\La^2u^4(\La^2-256u)\over 4(\La^4-256\La^2u+4096)^2}.
\ena
We set $y=\La^2/256u$ and $w=4y(1-y)$, then $z''=27w/(4w-1)^3$, 
 so we use the other cubic transformaion (\ref{eq:cub2})
and the quadratic transformation (\ref{eq:quad1}) subsequently, 
we get the expression for the massless case,
\bea
{\pa a\over \pa u}&=&{\sqrt 2\over 2}{1\over 2\sqrt u}F\left(
{1\over 2},{1\over 2};1;y\right),\nonumber \\
{\pa a_D\over \pa u}&=&i{\sqrt 2\over 2}{1\over 2\sqrt u}
\left[{2\ln 4-i\pi\over 2\pi}F\left({1\over 2},{1\over 2}
;1;y\right)-{1\over 2\pi}F^{*}\left({1\over 2},{1\over 2}
,1,y\right) \right].
\ena
Integrate with respect to $u$, we can recover the previous 
 result for $a$ and $a_D$ \cite{IY}. Expression in the strong coupling region can be obtained quite similarly. 

As the example of the theory which has the conformal points, 
we treat two cases where $\Delta$ become factorized multiple; 
one is the theory with  $m_1=m_2=m_3=\La/8$\cite{APSW,EHIY} and another one is 
$m_1=m_2=0,\ m_3=\La/16$ case. Of course other possibilities 
exist but we will not consider these possibilities for simplicity. 

In $m_1=m_2=m_3=\La/8$ case, $\Delta=0$ has a 4-fold root as
\bea
\Delta&=&-{\La^2\over 2^{20}}(32u-\La^2)^4(256u+19\La^2),\ \ 
D=-{(32u-\La^2)^2\over 64},\\
z''&=&-{27\over 16}{\La^2(256u+19\La^2)\over (32u-\La^2)^2}.
\ena
If we take
\bea
y={-27\La^2\over 256u-8\La^2},
\ena
then $z''=4y(1-y)$. Using the quadratic transformation (\ref{eq:quad1}) 
, we get ${\pa a\over \pa u}$ and ${\pa a_D\over \pa u}$
\bea
{\pa a\over \pa u}&=&{\sqrt2 \over 2}\left(-{27\La^2\over 256}\right)^
{-{1\over 2}}y^{1/2}
 F\left({1\over 6},
{5\over 6},1,y\right)\\
{\pa a_D\over \pa u}&=&i{\sqrt 2\over 2}
\left(-{27\La^2\over 256}\right)^{-{1\over 2}}
y^{1/2} \left[{\left(3\ln 3+2\ln 4-i\pi\right)\over 2\pi}
F\left({1\over 6},{5\over 6},1,y\right)\right.\\
& &\hspace{6cm}\left.-{1\over 2\pi}F^{*}
\left({1\over 6},{5\over 6},1,y\right)\right].\nonumber
\ena
Integrating with respect to $u$, we get $a$ and $a_D$ in the weak coupling 
region as
\bea
a&=&-{\sqrt2 \over 2}\left(-{27\La^2\over 256}\right)^{1\over 2}
{y^{-{1\over 2}}}
_3F_2\left({1\over 6},{5\over 6},-{1\over 2};
1,{1\over 2};y\right),\\
a_D&=&-i{\sqrt2 \over 2}\left(-{27\La^2\over 256}\right)^{1\over 2}
y^{-{1\over 2}}
\left[{(3\ln 3+2\ln 4-i\pi-4)\over 2\pi}
\, _3F_2\left({1\over 6},{5\over 6},-{1\over 2};
1,{1\over 2};y\right)\right.\\
& &\hspace{5cm}\left.-{1\over 2\pi}\, 
_3F^{*}_2\left({1\over 6},{5\over 6},-{1\over 2};
1,{1\over 2};y\right)\right].\nonumber
\ena
By using  the analytic continuation from the weak coupling region, we obtain the 
expression around the conformal point $u=\La^2/32$ as follows; 
\bea
a&=&-{\sqrt2 \over 2}\left(-{27\La^2\over 256}\right)^{1\over 2}
y^{-1/2}\left[{3\over 2}
{\Gamma({2\over 3})\over \Gamma({5\over 6})\Gamma({5\over 6})}
(-y)^{-1/6}\, _3F_2\left({1\over 6},{1\over 6},{2\over 3};{1\over 3}
,{5\over 3};{1\over y}\right)
\right. \nonumber \\
& &\ \ +\left.{3\over 4}{\Gamma(-{2\over 3})\over \Gamma({1\over 6})
\Gamma({1\over 6})}
(-y)^{-5/6}\, _3F_2\left({5\over 6},{5\over 6},{4\over 3};{5\over 3},
{7\over 3};{1\over y}\right)
\right],\\
a_D&=&-i{\sqrt 3\over 2}
{\sqrt2 \over 2}\left(-{27\La^2\over 256}\right)^{1\over 2}
y^{-1/2}\left[{3\over 2}
{\Gamma({2\over 3})\over \Gamma({5\over 6})\Gamma({5\over 6})}
(-y)^{-1/6}\, _3F_2\left({1\over 6},{1\over 6},{2\over 3};{1\over 3}
,{5\over 3};{1\over y}\right)
\right. \nonumber \\
& &\ \ -\left.{3\over 4}{\Gamma(-{2\over 3})\over \Gamma({1\over 6})
\Gamma({1\over 6})}
(-y)^{-5/6}\, _3F_2\left({5\over 6},{5\over 6},{4\over 3};{5\over 3}
,{7\over 3};{1\over y}\right)
\right].\ena

Next we consider $m_1=m_2=0,\ m_3=\La/16$ case. In this case  
$\Delta=0$ has one triple root and one double root as
\bea
\Delta&=&{\La^2\over 2^{27}}(\La^2-128u)^3(\La^2+128u)^2,\ \ 
D=-{(7\La^2-128u)(\La^2-128u)\over 1024},\\
z''&=&54{\La^2(\La^2+128u)^2\over (7\La^2-128u)^3}.
\ena
If we take
\bea
w={2\La^2\over 128u+\La^2},
\ena
then $z''=27w/(4w-1)^3$. So we use the cubic transformation (\ref{eq:cub2}), 
and use the identity (\ref{eq:ide}) ,
 we obtain ${\pa a\over \pa u}$ and ${\pa a_D\over \pa u}$
\bea
{\pa a\over \pa u}&=&{\sqrt2 \over 2}
\left(-{\La^2\over 64}\right)^{-{1\over 2}}y^{1/2}
F\left({1\over 4},
{3\over 4},1,y\right)\nonumber \\
{\pa a_D\over \pa u}&=&i{\sqrt 2\over 2}
\left(-{\La^2\over 64}\right)^{-{1\over 2}}
y^{1/2} \left[{\left(3\ln 4-i\pi\right)\over 2\pi}
F\left({1\over 4},{3\over 4},1,y\right)\right.\\
& &\hspace{6cm}\left.-{1\over 2\pi}F^{*}
\left({1\over 4},{3\over 4},1,y\right)\right],\nonumber
\ena
where
\bea
y={2\La^2\over -128u+\La^2}.
\ena
Integrate with respect to $u$, 
we obtain 
 $a$ and $a_D$ in the weak coupling region as
\bea
a&=&-{\sqrt 2\over 2}\left(-{\La^2\over 64}\right)^{1\over 2}{y^{-{1\over 2}}}
_3F_2\left({1\over 4},{3\over 4},-{1\over 2};1,{1\over 2};y\right),\\
a_D&=&-i{\sqrt 2\over 2}\left(-{\La^2\over 64}\right)^{1\over 2}y^{-{1\over 2}}
\left[{(3\ln 4-i\pi-4)\over 2\pi}\, _3F_2\left(
{1\over 4},{3\over 4},-{1\over 2};1,{1\over 2};y\right)\right.\\
& &\hspace{4cm}\left.-{1\over 2\pi}\, _3F^{*}_2\left(
{1\over 4},{3\over 4},-{1\over 2};1,{1\over 2};y\right)
\right].\nonumber
\ena
 Since this expression is the same as $N_f=2$ case (\ref{eq:aw})
 and (\ref{eq:aDw} except the argument $y$, 
we obtain the expression around the conformal point $u=\La^2/128$ 
by replacing the 
argument of (\ref{eq:paa}) and (\ref{eq:paaD}) to $y=2\La^2/(-128u+\La^2)$.

\ 

\sect{Summary}

We have derived a formula for the periods of
$N=2$ supersymmetric $SU(2)$ Yang-Mills theory with
massive hypermultiplets both in the weak coupling region and in the 
strong coupling region by using the identities of the hypergeometric 
functions. We also show how to deal with the theories with conformal 
points by using the formula.

The approach to evaluate the integral
 is useful when Picard-Fuchs equation is not solved
by any special functions.
Similar situation occurs when we consider the theories 
having higher rank gauge groups. In these case, we no longer 
expect that the similar transformations exist. We should know how to 
evaluate the dual pair of fields by another method,  which will be 
reported in a separate paper\cite{MS}.



\newpage
\begin{thebibliography}{99}
\bibitem{SW}
N. Seiberg and E. Witten, Nucl. Phys. B426 (1994) 19; 
Nucl. Phys. B431 (1994) 484.

\bibitem{KLTY}
A. Klemm, W. Lerche, S. Theisen and S. Yankielowics, Phys. Lett. 
B344 (1995) 196.\\
C. P. Argyres and A. Farragi, Phys. Rev. Lett. 74 (1995) 3931.


\bibitem{APS}
P. C. Argyres, R. Plesser and A. Shapere, Phys. Rev. Lett. 75 (1995) 1699;
 hep-th/9505100.\\
J. Minahan and D. Nemeshansky, Nucl. Phys. B464 (1996) 3;
 hep-th/9507032.

\bibitem{HO}
A. Hanany and Y. Oz, Nucl. Phys. B452 (1995) 283; hep-th/9505073.

\bibitem{DS}
U. Danielsson and B. Sundborg, Phys. Lett. B358 (1995) 273; hep-th/9504102.\\
A. Brandhuber and K. Landsteiner, Phys. Lett. B358 (1995) 73.

\bibitem{Hanany}
A. Hanany, Nucl. Phys. B466 (1996) 85; hep-th/9509176.

\bibitem{AS}
P.C.Argyres and A. Shapere, Nucl. Phys. B461 (1996) 437; hep-th/9509175.


\bibitem{CDF}
A.Ceresole, R. D'Auria and S. Ferrara, Phys. Lett. B339 (1994) 71.

\bibitem{KLT}
A. Klemm, W. Lerche and S. Theisen, Int. Jour. Mod. Phys. A11(1996) 1929;
 hep-th/9505015.

\bibitem{IY}
K. Ito and S.-K. Yang, Phys. Lett. B366 (1996) 165.

\bibitem{Matone}
M. Matone, Phys. Lett. B357(1995) 342.

\bibitem{Ohta}
Y. Ohta, hep-th/9604051, 9604059. 

\bibitem{HTF}
see for example, A.Erd\'{e}lyi et.al., ``\it Higher
Transcendal Fuctions'',\rm (McGraw-Hill, New York)
 Vol.\bf 1. \rm

\bibitem{APSW}
P.C. Argyres, R.N. Plesser, N.Seiberg and E. Witten, 
Nucl. Phys. B461 (1996) 71; hep-th/9511154.

\bibitem{EHIY}
T. Eguchi, K. Hori, K. Ito and S.K. Yang, 
Nucl. Phys. B471 (1996) 430; hep-th/9603002.

\bibitem{MS}
T. Masuda and H. Suzuki, hep-th/9609065.


\end{thebibliography}
\end{document}